\documentclass[12pt,twoside]{article}

\usepackage{palatino, geometry, url}

\usepackage[colorlinks=true,linkcolor=blue,citecolor=blue,urlcolor=blue]{hyperref}

\usepackage{cite}
\usepackage{footnote}
\usepackage{amsmath,amssymb,amsfonts}
\usepackage{graphicx}
\usepackage{array}
\usepackage{algorithmic}
\usepackage{algorithm}
\usepackage{array}
\usepackage[caption=false,font=normalsize,labelfont=sf,textfont=sf]{subfig}
\usepackage{longtable}
\usepackage{physics}
\usepackage{textcomp}
\usepackage{stfloats}
\usepackage{url}
\usepackage{verbatim}
\usepackage{graphicx}
\usepackage{mathtools}
\usepackage{subfig}
\usepackage{float}
\usepackage{textcomp}
\usepackage{amsfonts}
\usepackage{mathtools}
\usepackage{amssymb}
\usepackage{gensymb}
\allowdisplaybreaks

\usepackage{fancyhdr}

\begin{document}

\title{Adaptive Safe Reinforcement Learning-Enabled Optimization of Battery Fast-Charging Protocols}

\author{Myisha A. Chowdhury${}^\dag$, Saif S.S. Al-Wahaibi${}^{\dag}$, and Qiugang Lu${}^\dag$\thanks{Corresponding author: Qiugang Lu; Email: jay.lu@ttu.edu}\\
\\
{\small ${}^\dag$Department of Chemical Engineering, Texas Tech University, Lubbock, TX 79409, USA}
\\
}
\date{}
\maketitle

\begin{abstract}
Optimizing charging protocols is critical for reducing battery charging time and decelerating battery degradation in applications such as electric vehicles. Recently, reinforcement learning (RL) methods have been adopted for such purposes. However, RL-based methods may not ensure system (safety) constraints, which can cause irreversible damages to batteries and reduce their lifetime. To this end, this work proposes an adaptive and safe RL framework to optimize fast charging strategies while respecting safety constraints with a high probability. In our method, any unsafe action that the RL agent decides will be projected into a safety region by solving a constrained optimization problem. The safety region is constructed using adaptive Gaussian process (GP) models, consisting of static and dynamic GPs, that learn from online experience to adaptively account for any changes in battery dynamics. Simulation results show that our method can charge the batteries rapidly with constraint satisfaction under varying operating conditions.
\end{abstract}

{\bf Keywords}: Battery fast-charging; Safe reinforcement learning; Action projection; Gaussian process; Dynamic environment

\section{Introduction}
\label{sec:introduction}
The design of fast-charging strategies for Lithium-ion batteries is of significance to reduce the charging time, alleviate the mileage anxiety for electric vehicles (EV) \cite{liu2019brief}, and optimize EV fleet economics \cite{korkas2017adaptive}. Fast-charging of batteries with overly large currents may accelerate the growth of lithium dendrite and electrolyte decomposition, leading to degraded battery lifetime \cite{jana2019electrochemomechanics}. Moreover, the excessive heat generated from aggressive charging may bring significant safety hazards, including fires and explosions \cite{keyser2017enabling}. Thus, reasonable fast-charging protocols shall be deliberately designed to reduce the charging time while accounting for the safety and degradation of batteries \cite{wei2021deep}. 

Designing fast-charging strategies for batteries has been widely studied in the literature \cite{severson2019data,ahmed2017enabling}. Existing approaches can be broadly categorized into three groups: ad-hoc, model-based, and data-driven methods \cite{thakur2023state,bose2022study}. For ad-hoc methods, examples include the classical constant-current constant-voltage (CCCV) charging, pulse charging, and other variants \cite{notten2005boostcharging,purushothaman2006rapid}. These methods are simple in implementation but overly conservative for optimality \cite{jiang2022fast}. For model-based methods, the design of fast-charging protocols is often formulated as a constrained optimization with battery electrochemical models included in the constraints \cite{klein2011optimal}. However, such methods are restricted by the high complexity of solving the battery models \cite{zou2017electrochemical}. For data-driven methods, battery models are no longer needed, and the optimal charging profile is acquired by machine learning (ML) \cite{attia2020closed,pozzi2022deep} and deep reinforcement learning (DRL) \cite{hao2023adaptive}. Unlike other ML methods, e.g., those with Bayesian optimization \cite{attia2020closed}, DRL-based methods can \textit{adaptively} learn the optimal charging protocol as battery ages and parameter drifts, thus receiving wide attention. 

Reinforcement learning (RL) is a sequential decision-making method that iteratively interacts with the environment by observing the state, generating an action, and deploying the action to the environment \cite{sutton2018reinforcement}. The decision-making of the RL agent improves by learning from the environment reward to eventually optimize a prescribed objective. DRL-based fast-charging optimization is firstly studied in \cite{park2020reinforcement,park2022deep}, where the deep deterministic policy gradient (DDPG) method is employed for designing charging protocols that adapt to environmental changes. In \cite{chang2020control}, Q-learning is combined with recurrent neural network (RNN) to design charging profiles to minimize the cost of the participation of battery storage in the power grid. The authors in \cite{wei2021deep} study the integration of knowledge-based multiphysics constraints with DRL to balance between the charging speed and battery thermal constraints/degradation. Other advances include DRL-based optimal charging for extending battery lifetime \cite{kim2023optimal}, battery pack fast-charging with balance awareness \cite{yang2023balancing}, model-based DRL for fast-charging design \cite{hao2023adaptive}, and RL-based multi-stage constant current charging \cite{elouazzani4218801smart}. 

Despite the rapid development of DRL-based optimization of fast-charging strategies, the majority of reported methods cannot \textit{strictly} ensure constraint satisfaction throughout the charging process due to the usage of soft penalty for constraint violations in the reward \cite{wei2021deep}. This can lead to degraded battery life or safety risks. Given this knowledge gap, the proposed work will develop safe DRL-based battery fast-charging strategies to rapidly charge the battery while meeting the (safety) constraints with a high probability. Specifically, this work proposes to use probabilistic Gaussian process (GP) models online to approximate the constraint functions. With the trained and updated GP models, for any decided RL action (charging current) from the agent, the constraint values are predicted along with an uncertainty bound. Such information will be employed to construct a safety region, which is a set of safe actions that can satisfy all constraints. The agent can then project its actions into the safety region to ensure safety prior to deploying the action. The projection step is performed by solving a constrained optimization problem. 

In this work, the classical twin-delayed DDPG (TD3) algorithm will be used as backbone RL algorithm due to its deterministic nature for fast convergence and ability to mitigate the overestimation of Q values incurred with the DDPG algorithm \cite{fujimoto2018addressing}. The main contributions of this work are as follows: 

(1) GP-based models are presented to adaptively learn the constraint function. Safety regions of the action can then be constructed, and any unsafe action will be projected into this region before deployment to meet constraints. 

(2) Our work also proposes an adaptive safe RL-based optimization of battery fast-charging protocols. The optimized protocol can minimize the charging time while respecting the constraints with a high probability even under varying ambient and internal conditions. 
Note that how to ensure hard constraints of the system is a common issue for RL-based methods (not only for battery fast charging). Thus, our method is of significance for applications beyond batteries where safety constraints are critical and also the system dynamics vary over operating conditions.

The remainder of this paper is organized as follows. Section \ref{sec:preliminaries} presents the fundamentals of RL, including a brief introduction to the TD3 algorithm. Section \ref{sec:methodology} demonstrates the proposed adaptive safe RL method, and details are given about the GP surrogate modeling and action projection. The specific adaptive safe RL-based battery fast-charging optimization is given in Section \ref{sec:battery_optimization}. The effectiveness of the proposed method is validated on the PyBaMM simulator in Section \ref{sec:result}. The conclusions are given in Section \ref{sec:conclusion}. 

\section{Preliminaries}
\raggedbottom
\label{sec:preliminaries}
This section will give a brief overview of RL and the TD3 algorithm that will be implemented in this work. A detailed description of the RL algorithms can be found in \cite{sutton2018reinforcement}.

\subsection{Reinforcement learning}
\label{subsec: reinforcement_learning}
A general RL framework consists of an agent and an environment. The agent is a continuously learning decision-maker, while the environment refers to all the system elements that the agent interacts with. The decision-making process of RL can be formalized as a Markov decision process (MDP), $\mathcal{M}: (\mathcal{S}, \mathcal{A}, \mathcal{P}, \gamma, r)$, where $\mathcal{S}$ and $\mathcal{A}$ are the state and action space, respectively, $\mathcal{P}:\mathcal{S}\times \mathcal{A} \times \mathcal{S}\rightarrow \left[0,1\right]$ is the state transition probability, $\gamma \in [0,1]$ is the discount factor, and $r: \mathcal{S} \times \mathcal{A} \rightarrow \mathbb{R}$ is the reward function. In a MDP setting, at each time step $t$, the agent observes the state vector $s_t \in \mathcal{S}$, generates an action $a_t \in \mathcal{A}$ and deploy it to the environment, which results in one-step environment evolution to $s_{t+1} \in \mathcal{S}$ based on the transition probability $\mathcal{P}(s_{t+1}|s_t,a_t)$. At any time step $t$, the policy $\pi(a_t|s_t):s_t\rightarrow \mathcal{P}(a_t)$ gives the probability of selecting action $a_t$ given the environment state $s_t$. Additionally, each state is evaluated by the state-value function, $V^\pi: \mathcal{S} \rightarrow \mathbb{R}$, which assigns a value to each state by calculating the expected total cumulative reward, starting from that state and following policy $\pi$ henceforth
\begin{equation}
V^\pi (s_t) = \mathbb{E}_\pi\left[\sum\nolimits_{k=0}^{\infty}\gamma^k r(s_{t+k},a_{t+k})\right].\label{eq:valuefunction}
\end{equation} 
where $\gamma$ is the discount factor. Moreover, the state-action value function $Q^{\pi}:\mathcal{S} \times \mathcal{A} \rightarrow \mathbb{R}$, also known as Q-function, evaluates each state-action pair by calculating the expected total cumulative reward starting from a state and the selected action by policy $\pi$ based on that state:
\begin{equation}
Q^\pi (s_t,a_t) = \mathbb{E}_\pi\left[\sum\nolimits_{k=0}^{\infty}\gamma^k r(s_{t+k},a_{t+k})|s_t,a_t\right]. \label{eq:qvaluefunction}
\end{equation} 
One can re-write \eqref{eq:qvaluefunction} recursively with the Bellman equation   
\begin{align}
Q^{\pi}(s_t, a_t)=&\mathbb{E}_{s_{t+1}\sim p(\cdot|s_t,a_t)}[ r(s_t,a_t) + \gamma \mathbb{E}_{a_{t+1}\sim \pi(\cdot|s_{t+1})}\left[Q^{\pi}(s_{t+1}, a_{t+1})\right]], \label{eq:bellmanequation}
\end{align}
where $p(\cdot|s_t,a_t)$ is the transition probability between successive states of the environment, provided that the current state and action are $s_t$ and $a_t$, respectively. The objective of the MDP is to find the optimal policy $\pi^\ast$ that can yield the maximal Q-value $Q^\ast:=\max_\pi ~Q^\pi(s_t,a_t)$, that is,
\begin{equation}
\pi^{*}=\arg\max_{\pi}~Q^\pi(s_t,a_t). \label{eq: optimal_q_value}
\end{equation}

\subsection{TD3 algorithm}
\label{subsec:td3}
Similar to the DDPG algorithm \cite{lillicrap2015continuous}, the TD3 method is also a deterministic actor-critic method for continuous state-action space \cite{fujimoto2018addressing}. Traditional DDPG method suffers from the overestimation bias of the Q value due to the repetitive maximization operation over the target Q networks (critics) in calculating the temporal difference (TD) target. Such bias leads to inaccurate estimations of the Q value. To address this issue, the TD3 adopts two target critic networks, and the smaller one is used to compute the TD target \cite{fujimoto2018addressing}
\begin{equation}
y_t=r(s_t,a_t)+\gamma \min_{i=1,2}Q_{\theta^{\prime}_i}(s_{t+1},\mu_{\phi^{\prime}}(s_{t+1})),
\end{equation}
where $\theta^{\prime}$ and $\phi^{\prime}$ are the parameters of the target critic network $Q_{\theta^{\prime}_i}$ and target actor network $\mu_{\phi^{\prime}}$, and $r(s_t,a_t)$ is the reward. The TD3 also has two critic networks $Q_{\theta_i}(s_t,a_t)$, whose parameters $\theta_i,i=1,2$, are updated via minimizing the TD error by gradient descent
\begin{equation}
\nabla_{\theta_i}\frac{1}{|B|}\sum_{(s_t, a_t, r_t, s_{t+1})\in B} \left[y_t-Q_{\theta_i}(s_t,a_t) \right]^2,
\end{equation}
where the batch $B$ contains transition tuples $(s_t,a_t,r_t,s_{t+1})$ sampled from the reply buffer. The actor-network $\mu_{\phi}(s_t)$ approximates the policy mapping from state $s_t$ to action $a_t$, with parameters $\phi$ whose update is via gradient ascent  
\begin{equation}
\nabla_{\phi}\frac{1}{|B|}\sum_{s_t\in B} Q_{\theta_1}(s_t,\mu_{\phi}(s_t)).
\end{equation}
The update of target network parameters follows that of the Polyak averaging ($0<\rho<1$, e.g., $\rho=0.99$):
\begin{equation}
\theta_{i}^{\prime}\leftarrow \rho\theta_{i}^{\prime}+(1-\rho)\theta_{i},~~\phi^{\prime}\leftarrow \rho\phi^{\prime}+(1-\rho)\phi.
\end{equation}

\section{Action projection-based safe RL}
\label{sec:methodology}
\subsection{Gaussian process (GP) model}
This section first introduces the basics of GP models, and they will be employed in our safe RL for building surrogate models of the constraint functions. As a non-parametric method \cite{wang2020intuitive}, GP assumes that the function to be learned $q(x)$ is sampled from a Gaussian distribution with the mean function $\mathbb{E}\left[q(x)\right]=m(x)$ (often assumed to be zero) and covariance function $cov[q(x),q(x^\prime)]=k(x,x^\prime)$, where $k(\cdot,\cdot)$ is the kernel function. The covariance function defines the output correlation of the function $q(\cdot)$ at inputs $x$ and $x^\prime$. One commonly used kernel function due to its smooth and differentiable form is the radial-basis function \cite{wang2020intuitive},
\begin{equation}
k(x,x^\prime)=\sigma_f^2 \exp \left[-\|x-x^\prime\|^2/{2l^2}\right], \label{eq:RBF}
\end{equation}
where $\sigma_f^2$ and $l^2$ control the vertical span of the curve and the correlation drop speed as distance increases, respectively. Consider $n$ sampled input-output pairs $\{x_{1:n},y_{1:n}\}$, where $y_i=q(x_i)$. With GP, for a query point $x^{\ast}$, the function value $y^\ast=q(x^\ast)$ is inferred by conditional multivariate Gaussian distribution $\mathcal{N}(\cdot,\cdot)$ \cite{richardson2017gaussian}:
\begin{align}
& 	p(y^\ast|x^\ast,x_{1:n},y_{1:n}) = \mathcal{N}(y^\ast|\mu_{n}, \sigma_n), \label{eq:gp_model} \\
& \mu_n:=\mu(x^\ast,x_{1:n},y_{1:n})= K(x^\ast,x_{1:n}) K(x_{1:n},x_{1:n})^{-1} y_{1:n},\nonumber \\
& \sigma_n :=\sigma(x^\ast,x_{1:n})= K(x^\ast,x^\ast) - K(x^\ast,x_{1:n}) K(x_{1:n},x_{1:n})^{-1} K(x_{1:n},x^\ast), \nonumber
\end{align}
where $\mu_n$ and $\sigma_n$ are the posterior mean and variance, and $K$ is the covariance matrix constructed from the kernel function $k(\cdot,\cdot)$ with entry $K_{i,j}=k(x_i,x_j)$.

\subsection{Action projection-based safe RL}
\begin{figure}[tbh]
\centerline{\includegraphics[width=0.45 \columnwidth, height=0.4 \columnwidth]{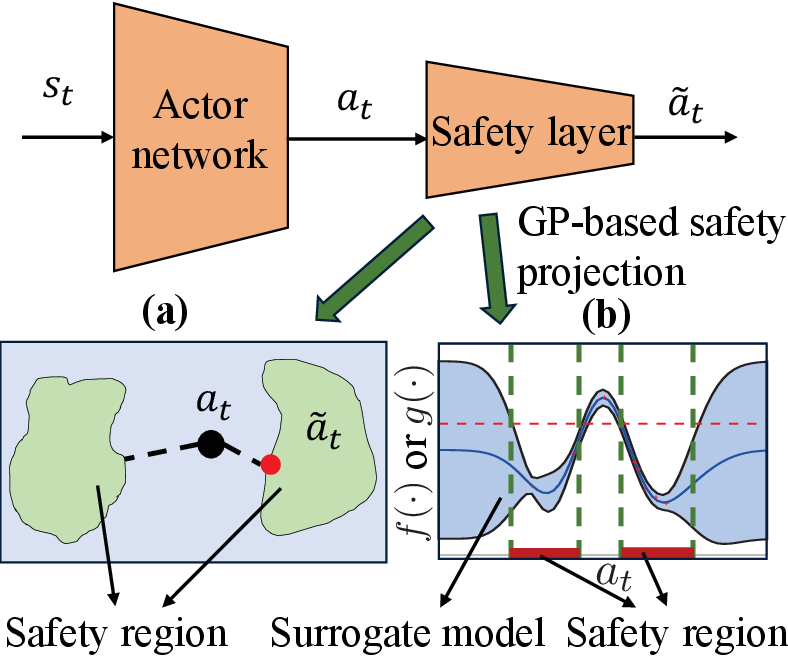}}
\caption{Schematics of the proposed action projection-based safe RL.}
\label{fig:gp}
\end{figure}

Although RL has been increasingly employed for sequential decision-making problems with black-box environments, its lack of ability to ensure constraints hinders the usage of RL in real-world safety-critical applications. To this end, this work proposes an action projection method (see Fig. \ref{fig:gp}) where an additional safety layer is appended after the actor network. The action projection below will be solved by the safety layer: 
\begin{align}
\min_{\tilde{a}_t\in\mathcal{A}} ~~||a_t-\tilde{a}_t||^2_2,~~s.t., \quad z_{t+1}=q(z_{t},h_t,\tilde{a}_t) \in \mathcal{Z}, \label{eq:general_optimization}
\end{align}
where $a_t$ is the raw agent action, $\mathcal{A}$ is the action space, and $\tilde{a}_t$ is the (projected) closest point to $a_t$ in the safety region (green area in Fig. \ref{fig:gp} (a)), defined by next-step \textit{safety variable} $z_{t+1}$ and feasible set $\mathcal{Z}$. In general, constraint dynamics $z_{t+1}=q(z_{t},h_t,\tilde{a}_t)$ is \textit{black-box}, e.g., it can the (unknown) relation between next-step battery temperature and the charging current to be applied ($h_t$ are some known variables at $t$). Thus, it is proposed to use GP models to build data-driven surrogates of $q(\cdot)$ from historical data. Then the prediction (posterior mean and variance) of $z_{t+1}$ from the GP model \eqref{eq:gp_model} can be utilized to replace the constraint in \eqref{eq:general_optimization}. Fig. \ref{fig:gp} (b) gives an example where the posterior mean (blue curve) and posterior variance (blue shaded area) of $z_{t+1}$, inferred by the trained GP model, are used to construct the safety region (red bars) where $\mathcal{Z}$ is a threshold (red dashed) on the upper uncertainty bound of the posterior prediction. 

\section{Safe RL for fast charging optimization}
\label{sec:battery_optimization}
This section details the proposed adaptive safe RL method for optimizing battery fast-charging protocols while respecting system constraints with a high probability. For demonstration, the TD3 algorithm is used as the backbone RL method, which can be easily extended to other RL algorithms, such as DDPG \cite{lillicrap2015continuous}, SAC \cite{haarnoja2018soft}, etc. Note that the proposed safe and adaptive RL method falls into the online off-policy category (similar to TD3, DDPG, etc.) since the agent needs to interact online with the environment to gather new data and also access history data (collected under a different policy) for the training \cite{levine2020offline}.

\subsection{Fast-charging optimization formulation}
This work aims to develop an RL-based battery fast-charging protocol while respecting system constraints. The fast-charging optimization problem can be formulated as \cite{hao2023adaptive}
\begin{align}
&\max_{\underline{I} \leq I_t \leq \bar{I}} \quad-t_f,\nonumber\\ 
&s.t.  \quad T_t \leq \bar{T}, \quad V_t \leq \bar{V}, \nonumber\\
& \quad \quad SOC(t_0) = SOC_0, \quad SOC(t_f)=SOC_{ref}, \label{eq:battery_optimization_general}
\end{align}
where $t_f$ and $I_t$ is the charging duration and current, respectively. Additionally, $T_t$ and $V_t$ are the temperature and voltage of the battery at time $t$, respectively, $SOC_0$ is the initial state-of-charge (SOC) at time $t_0$, and $SOC_{ref}$ is the SOC at time $t_f$ when considering the charging cycle is complete. Finally, $\bar{T}$ and $\bar{V}$ are the maximum allowable operating temperature and voltage, respectively. 

\subsection{Static safe RL for fast charging optimization}
\label{subsec:fixed_gp}
This section presents our safe RL framework for battery fast-charging optimization. The RL environment is the battery, and the action $a_t$ is the charging current $I_t$ to be decided. The safety projection for problem \eqref{eq:battery_optimization_general} can be formulated as
\begin{align}
&\min_{\tilde{a}_t\in\mathcal{A}}~~ ||a_t-\tilde{a}_t||^2_2,\label{eq:safe_gp_rl_obj} \\ s.t. \quad &T_{t+1}=f(T_{t},\tilde{a}_{t-1},\tilde{a}_t)  \leq \bar{T}, \label{eq:safe_gp_rl_c1} \\
& V_{t+1}=g(V_{t},\tilde{a}_{t-1},\tilde{a}_t)  \leq \bar{V},   \label{eq:safe_gp_rl_c2}
\end{align}
where black-box functions $f(\cdot)$ and $g(\cdot)$ predict next-step temperature and voltage respective for the given current $\tilde{a}_t$. Static GP models $\hat{f}_s(\cdot)$ and $\hat{g}_s(\cdot)$ are trained with historical data to approximate above constraint functions. The data of the first $M$ RL training episodes ($n$ samples) are selected as the historical data: 
$\{x^{f}_{1:n-1},T_{2:n}\}$ where $x^{f}_i=[T_{i}, a_{i-1}, a_i]^\top$, and $\{x^{g}_{1:n-1},V_{2:n}\}$, where $x^{g}_i=[V_{i}, a_{i-1}, a_i]^\top$. Note that the training data for static GP is collected prior to the deployment of action projection. The L-BFGS algorithm is used to train these two GP models. Starting from episode $M+1$, for any query RL action $\tilde{a}_t$ at time $t$, the predicted posteriors by trained static GPs $\hat{f}_{s}(\cdot)$, $\hat{g}_{s}(\cdot)$, are  $p(T_{t+1}|\tilde{a}_t)\sim\mathcal{N}(\mu_{t}^{f},\sigma_{t}^{f})$ for temperature and $p(V_{t+1}|\tilde{a}_t)\sim\mathcal{N}(\mu_{t}^{g},\sigma_{t}^{g})$ for voltage. As in \eqref{eq:gp_model}, here $\mu_{t}^f=\mu(x_t^f,x_{1:n-1}^f,T_{2:n})$ and $\sigma_{t}^f=\sigma(x_t^f,x_{1:n-1}^f)$. Also, $\mu_{t}^g=\mu(x_t^g,x_{1:n-1}^g,V_{2:n})$ and $\sigma_{t}^g=\sigma(x_t^g,x_{1:n-1}^g)$, where $x_{t}^{f}=[T_{t},\tilde{a}_{t-1},\tilde{a}_t]$ contains present temperature $T_t$ and the previous projected action $\tilde{a}_{t-1}$, and also the query $\tilde{a}_t$ to be solved. Similarly, $x_{t}^{g}=[V_{t},\tilde{a}_{t-1},\tilde{a}_t]$.

With the predicted posteriors, one can approximate \eqref{eq:safe_gp_rl_c1}-\eqref{eq:safe_gp_rl_c2} by using the upper uncertainty bound (UUB) for our surrogate models. Then the problem \eqref{eq:safe_gp_rl_obj}-\eqref{eq:safe_gp_rl_c2} is cast as:
\begin{align}
&\min_{\tilde{a}_t\in\mathcal{A}} ~~~||a_t-\tilde{a}_t||^2_2,\label{eq:safe_gp0_obj}  \\ s.t. \quad
&\mu_t^f+\kappa \sigma_t^f\le \bar{T},~\mu_t^g+\kappa \sigma_t^g\le \bar{V}, \label{eq:model_gp0_vol}
\end{align}
where $\kappa$ is a hyper-parameter. Fig. \ref{fig:gp} (b) shows the $\pm 3 $ standard deviation uncertainty bound (blue shaded) constructed from \eqref{eq:model_gp0_vol} where the upper bound is used to construct the feasible region (red bars). For this method, once GP models are trained, they will be fixed for future RL episodes, thereby known as \textit{static GP-based safe RL}, with pseudo-code in Algorithm 1. 


\begin{longtable}{ll}

\hline
\multicolumn{2}{l}{\textbf{Algorithm 1: Static GP-based safe RL (based on TD3 algorithm)}} 
\\ \hline
1: & \textbf{Input:} initialize Q-function parameters $\phi_1$, $\phi_2$, and policy parameter $\theta$, empty \\
&  replay buffer $\mathcal{D}$ and training data storage $\mathcal{X}$; \\	
2: & Initialize target networks $\phi_1^{\prime}\leftarrow\phi_1$, $\phi_2^{\prime}\leftarrow\phi_2$, $\theta^{\prime}\leftarrow\theta$; \\
3: &for episode $m = 1$ to $M$, repeat  \\
4: & \quad \quad $t \leftarrow 1$; \\
5: & \quad \quad Observe state $s_t$ and select $a_t$ randomly in $\mathcal{A}$;\\
6:& \quad \quad   Execute $a_t$ in the environment; \\
7:& \quad \quad  Observe the next state $s_{t+1}$ and compute the reward $r_t$;\\
8:& \quad \quad  Store $\left(z_t, a_{t-1}, a_t\right)$ and $(z_{t+1})$ in $\mathcal{X}$ to train GP models, where $z_t$ is the \\& \quad \quad  variable of interest (e.g., $T_{t}$ and $V_{t}$ in \eqref{eq:safe_gp_rl_c1} and  \eqref{eq:safe_gp_rl_c2}) in the constraints; \\
9:& \quad \quad  Store the transition $\left(s_t,a_t,r_t,s_{t+1}\right)$ into the replay buffer $\mathcal{D}$; \\
10:& \quad \quad \textbf{Update RL agent parameters:} \\
11:& \quad \quad   Randomly select a batch  $B=\left\{\left(s_t,a_t,r_t,s_{t+1}\right)\right\}$  from $\mathcal{D}$ with \\ 
& \quad \quad  the  number of transitions as $|B|$; \\
12:& \quad \quad  Update the critic network parameters:\\
13: & \quad \quad \quad \quad  Compute the target for the Q-functions \\ 
& \quad \quad \quad \quad  $y_t=r_t + \gamma (\min_{i=1,2} Q_{\phi_i ^\prime}(s_{t+1},a_{t+1}(s_{t+1}))$; \\
14:& \quad \quad \quad \quad   Update the Q-function using cost function w.r.t. $\phi_i$: \\
& \quad \quad \quad \quad  $\nabla_{\phi_i}\frac{1}{|B|} \sum\limits_{\left(s_t,{a}_t,r_t,s_{t+1}\right)\in B}\left(Q_{\phi_{i}}(s_t,{a}_t) - y_t \right)^2$, $i = 1, 2$; \\
15:& \quad \quad  Update the actor and target networks:\\
16: &  \quad \quad \quad \quad   if $t \quad \% \quad d = 0$, where $d$ is the delayed update period\\
17: &  \quad \quad \quad \quad  \quad  \quad $\nabla_{\theta}\frac{1}{|B|}\sum \limits_{s_t \in B}Q_{\phi_1}(s_t, \mu_{\theta}(s_t))$,\\
18: &  \quad \quad \quad\quad \quad \quad $\phi_{i}^\prime\leftarrow \rho \phi_{i}^\prime+(1-\rho)\phi_{i}$ , $i = 1, 2$,\\
19:& \quad \quad\quad \quad \quad  \quad $\theta^\prime\leftarrow \rho\theta^\prime+(1-\rho)\theta$;\\
20: & \quad \quad  $t \leftarrow t+1$, until $s_{t+1}$ is terminal; \\
21:&  end for \\
22:& Train the GP models $\hat{f}_s(\cdot)$, $\hat{g}_{s}(\cdot)$ using $\mathcal{X}$ with L-BFGS; \\
23: &for episode $m = M+1$ to END, repeat  \\
24: & \quad \quad $t \leftarrow 1$ \\
25: & \quad \quad  Observe state $s_t$ and use actor network $\mu_{\theta}(\cdot)$ for action $a_t$; \\
26:& \quad \quad Predict the posterior mean and variance of next-step variable  \\
& \quad \quad with the trained static GPs $\hat{f}_s(\cdot)$, $\hat{g}_{s}(\cdot)$;\\
27: & \quad \quad  Solve the optimization problem in \eqref{eq:safe_gp0_obj}-\eqref{eq:model_gp0_vol} to find $\tilde{a}_t$;  \\
28:& \quad \quad Execute $\tilde{a}_t$, observe  $s_{t+1}$, and compute $r_t$;\\
29:& \quad \quad  Store the transition $\left(s_t, \tilde{a}_t, r_t, s_{t+1}\right)$ into $\mathcal{D}$; \\
30:& \quad \quad  Update the actor and critic  following steps 10-19; \\
31: & \quad \quad $t \leftarrow t+1$, until $s_{t+1}$ is terminal or $m$ reaches END; \\
32:& end for \\
\hline
\end{longtable}

\subsection{Adaptive safe RL for varying environments}
\label{subsec:adaptive_gp}
The previously proposed RL algorithm with a GP-based safety layer assumes a static environment. However, the GP models trained with offline data cannot account for the changes in environments and system dynamics, and hence may not adaptively predict future values in such situations. Specific to batteries, changes in the operating conditions such as ambient temperature or battery aging alter the internal dynamics of the battery \cite{xiong2020lithium}. Thus, it is essential to develop adaptive GP models that capture battery dynamics in real-time to ensure the effectiveness of action projection in the case of changes in ambient conditions or battery aging.

Our method will employ the previous static GP models trained with the data from the first $M$ RL episodes as baseline models to capture the \textit{overall trends} of variables to be predicted. Then the differences between true variable values and posterior means $\mu_t^f$ and $\mu_t^g$ from the static GPs $\hat{f}_s(\cdot)$ and $\hat{g}_s(\cdot)$, known as \textit{residuals}, are
\begin{align}
\Delta T_{t+1}:=T_{t+1}-\mu_t^{f}=f_{\Delta}(T_{t},\tilde{a}_{t-1},\tilde{a}_t), \label{eq:delta_f}\\
\Delta V_{t+1}:=V_{t+1}-\mu_t^{g}=g_{\Delta}(V_{t},\tilde{a}_{t-1},\tilde{a}_t), \label{eq:delta_g}
\end{align}
with black-box functions $f_{\Delta}(\cdot)$ and $g_{\Delta}(\cdot)$. It is proposed to use two GP models, known as dynamic GPs, $\hat{f}_\Delta(\cdot)$ and $\hat{g}_\Delta(\cdot)$, to capture such residuals. Such dynamic GPs will be updated at \textit{every} timestep throughout an RL episode and \textit{re-trained from scratch} at the beginning of a new episode. Consider the present episode $k>M$, and the current timestep $t$. The collected data up to $t$ for the present episode are $\{x_{1:t-1}^{f},\Delta T_{2:t}\}$ and $\{x_{1:t-1}^{g},\Delta V_{2:t}\}$, with $x_{i}^f=[T_{i},\tilde{a}_{i-1},\tilde{a}_i]^\top$ and $x_{i}^g=[V_{i},\tilde{a}_{i-1},\tilde{a}_i]^\top$. The two dynamic GPs are then trained based on these data. For time $t$, the posterior predictions are 
\begin{equation}
p(\Delta T_{t+1}|\tilde{a}_t)\sim \mathcal{N}(\bar{\mu}_{t}^f,\bar{\sigma}_t^f), ~p(\Delta V_{t+1}|\tilde{a}_t)\sim \mathcal{N}(\bar{\mu}_{t}^g,\bar{\sigma}_t^g),  \label{eq:posterior_delta_gp}
\end{equation}
where $\bar{\mu}_{t}^f=\mu(x_t^f,x_{1:t-1}^f,\Delta T_{2:t})$ and $\bar{\sigma}_t^f=\sigma(x_t^f,x_{1:t-1}^f)$. $\bar{\mu}_{t}^g$ and $\bar{\sigma}_{t}^g$ are defined analogously and are omitted here. With the posterior distributions for $\Delta T_{t+1}$ and $\Delta V_{t+1}$, the overall posterior predictions from the \textit{adaptive} GPs, consisting static and dynamic GPs, are shown to be
\begin{equation}
p(T_{t+1}|\tilde{a}_t)\sim\mathcal{N}(\hat{\mu}_t^f,\hat{\sigma}_t^f),~p(V_{t+1}|\tilde{a}_t)\sim\mathcal{N}(\hat{\mu}_t^g,\hat{\sigma}_t^g), \label{eq:adaptive_gp_prediction}
\end{equation}
where $\hat{\mu}_t^f=\mu_t^f+\bar{\mu}_t^f$ and $\hat{\mu}_t^g=\mu_t^g+\bar{\mu}_t^g$. The definitions of posterior variance $\hat{\sigma}_t^f$ and $\hat{\sigma}_t^g$ can be flexible, either the sum or other combinations of those from static or dynamic GPs. Our work chooses that of the static GP as the posterior variance of the adaptive GP, i.e., $\hat{\sigma}_t^f={\sigma}_t^f$ and $\hat{\sigma}_t^g={\sigma}_t^g$, since typically the data quality for identifying dynamic GP is low due to lack of excitation and non-uniform scattering of applied currents $\tilde{a}_{1:t-1}$ that often cause $\hat{\sigma}_t^f$ and $\hat{\sigma}_t^f$ to be excessively large. 

With adaptive GP surrogates of constraint functions, the original problem \eqref{eq:safe_gp_rl_obj}-\eqref{eq:safe_gp_rl_c2} can be approximated as 
\begin{align}
&\min_{\tilde{a}_t\in\mathcal{A}}~~||a_t-\tilde{a}_t||^2_2, \label{eq:adaptive_gp_obj}\\
s.t. \quad & (\mu_t^f+\bar{\mu}_t^f) + \kappa \sigma_t^f  \leq \bar{T},\label{eq:adaptive_gp_c1}\\
&(\mu_t^g+\bar{\mu}_t^g) + \kappa \sigma_t^g  \leq \bar{V}. \label{eq:adaptive_gp_c2}
\end{align}
As before, the quantity of training data for dynamic GP is limited since only the data from current episode is used. Additionally, non-uniform data acquired during some episodes, particularly if they are sparse in certain regions, can lead to large posterior uncertainties in those regions. This is because the GP has less information to make predictions in those areas. In contrast, the training data for static GP are more extensive and diverse than the dynamic GP. Therefore, the static GP often has a more informative confidence interval. 
\begin{longtable}{ll}
\hline
\multicolumn{2}{l}{\textbf{Algorithm 2: Adaptive GP-based safe RL (based on TD3 algorithm) }} 
\\ \hline
1: & Follow (1)-(22) in Algorithm 1 to train static GPs $\hat{f}_s(\cdot)$, $\hat{g}_{s}(\cdot)$;\\	
2: &for episode $m = M+1$ to END, repeat\\
3: &\quad \quad t $\leftarrow$ 1;\\
4: &\quad \quad Empty the data storage $\mathcal{R}$ for training dynamic GPs;\\
5: &\quad \quad for $t = 1$ to ending step of charging for episode $m$, repeat\\
6: & \quad \quad \quad \quad Observe state $s_t$ and use actor network $\mu_{\theta}(\cdot)$ for action $a_t$;\\
7: & \quad \quad \quad \quad if $t \le n$, where $n$ is the timestep starting action projection:\\
8:& \quad \quad \quad \quad \quad \quad  Execute $a_t$, observe the next state $s_{t+1}$, and reward $r_t$; \\
9:& \quad \quad\quad \quad else:\\
10:& \quad \quad\quad \quad \quad \quad Train the dynamic GP models with $\mathcal{R}$;\\
11:& \quad \quad\quad \quad \quad \quad Get adaptive GPs from static and dynamic GPs;\\
12:& \quad \quad \quad \quad\quad \quad Predict $z_{t+1}$ using \eqref{eq:adaptive_gp_prediction}; \\
13: & \quad \quad \quad \quad\quad \quad  Solve the optimization  in \eqref{eq:adaptive_gp_obj}-\eqref{eq:adaptive_gp_c2}, to find $\tilde{a}_t$; \\
14:& \quad \quad\quad \quad\quad \quad Execute $\tilde{a}_t$, observe  $s_{t+1}$, and compute $r_t$;\\
15:& \quad \quad \quad \quad Predict the variable of interest $\tilde{z}_{t+1}$ with static GP;\\
16:& \quad \quad\quad \quad Calculate the residual, $\Delta_{t+1} = z_{t+1}-\tilde{z}_{t+1}$;\\
17:& \quad \quad\quad \quad Store $[z_{t+1},a_{t-1},a_t]$ and $[\Delta_{t+1}]$ into $\mathcal{R}$;\\
18:& \quad \quad\quad \quad Store the transition $\left(s_t,\tilde{a}_t,r_t,s_{t+1}\right)$ into $\mathcal{D}$; \\
19:& \quad \quad\quad \quad Update the networks using 10-19 in Algorithm 1; \\
20: & \quad \quad \quad \quad $t \leftarrow t+1$, until $s_{t+1}$ is terminal; \\
21:&\quad \quad end for \\
22:& end for \\
\hline
\end{longtable}

\section{Simulation Results and Discussions}
\label{sec:result}
This section uses two case studies, one with fixed and the other with varying operating conditions, to validate the effectiveness of the proposed methods. For these studies, the PyBaMM \cite{sulzer2021python}, an open-source Python package, is used to perform high-fidelity simulation of battery dynamics. 

\subsection{Safe RL-based fast charging with fixed conditions}
This section evaluates the proposed action projection-based safe TD3 approach for optimizing battery fast-charging protocol. It is compared with the traditional TD3 algorithm without the added safety layer and the classical CCCV method. Note that here the safe TD3 based on static GP in Section \ref{subsec:fixed_gp} is referred to as ``safe TD3'' and the one based on adaptive GP in Section \ref{subsec:adaptive_gp} as ``adaptive safe TD3''.

This study aims to optimize the charging policy to charge the battery from $10\%$ to $80\%$ state-of-charge (SOC) in the shortest possible time while ensuring that the temperature and voltage remain within operational constraints. The maximum allowable temperature $\bar{T}$ and voltage $\bar{V}$ for this case are $45^{\degree} C$ and $4.3V$, respectively. Additionally, the ambient temperature is kept constant at $25^{\degree} C$ throughout the experiment. The charging current $a_t\in[0.05C, 4.5C]$, where $C$ is the charging rate. The sampling time is 10 seconds. For our experiment, the environment state consists of SOC, voltage, temperature, and the current applied at the previous time step: $s_t=[SOC_t, V_t, T_t, a_{t-1}]^\top$. At each timestep in an episode, the actor network is given the state $s_t$ as input and outputs a charging current $a_t$. The actor network has two hidden layers, each consisting of 128 nodes. The ReLU is used as the activation function and ADAM as the optimizer. The critic network receives the state $s_t$ and action $a_t$ as inputs and generates Q value to assess the policy. The critic network has a similar architecture to the actor network. The hyper-parameters of the proposed adaptive safe TD3 algorithm are fine-tuned by the trial-and-error method and listed in Table \ref{table: Hyperparameters}. 
\begin{table}[htbp]
\centering
\caption{List of hyperparameters used by the adaptive safe TD3.}
\label{table: Hyperparameters}
\begin{tabular}{llll}
\hline
Parameters & Values (Fixed condition) & Values (Varying condition)\\ \hline
Learning rate $\alpha$ for actors & 0.0005 & 0.00005\\
Learning rate $\beta$ for critics & 0.005 & 0.0005 \\
Batch size $B$ & 64 & 64 \\
Discount factor $\gamma$ in the return & 0.99 & 0.99 \\
Polyak averaging coefficient $\rho$ & 0.006 & 0.006\\
Initial noise variance $\sigma^2$  & 0.3 & 0.3\\
Noise decay factor & 0.025 & 0.025 \\ 
Length scale (RBF) & 1.0 & 1.0 \\
Noise level (white kernel) & $10^{-5}$ & $10^{-5}$\\		\hline 
\end{tabular}%
\end{table}

The reward function is designed to penalize the charging time while respecting the temperature and voltage constraints \cite{hao2023adaptive}, defined as 
\begin{equation}
r_t = -1 + \lambda_1 [V_t - \bar{V}]^{+} + \lambda_2 [T_t - \bar{T}]^{+}, \label{eq:reward}
\end{equation}
where the first term is the penalty on the charging time (steps). The second and third terms penalize the voltage and temperature constraint violations. The clip function $[\cdot]^+$ returns zero if constraints are satisfied, or the amount of violations otherwise. The weights are selected as $\lambda_1=15$, $\lambda_2=20$. 

\begin{figure}[!t]
\centerline{\includegraphics[width=0.9 \columnwidth, height=0.35 \columnwidth]{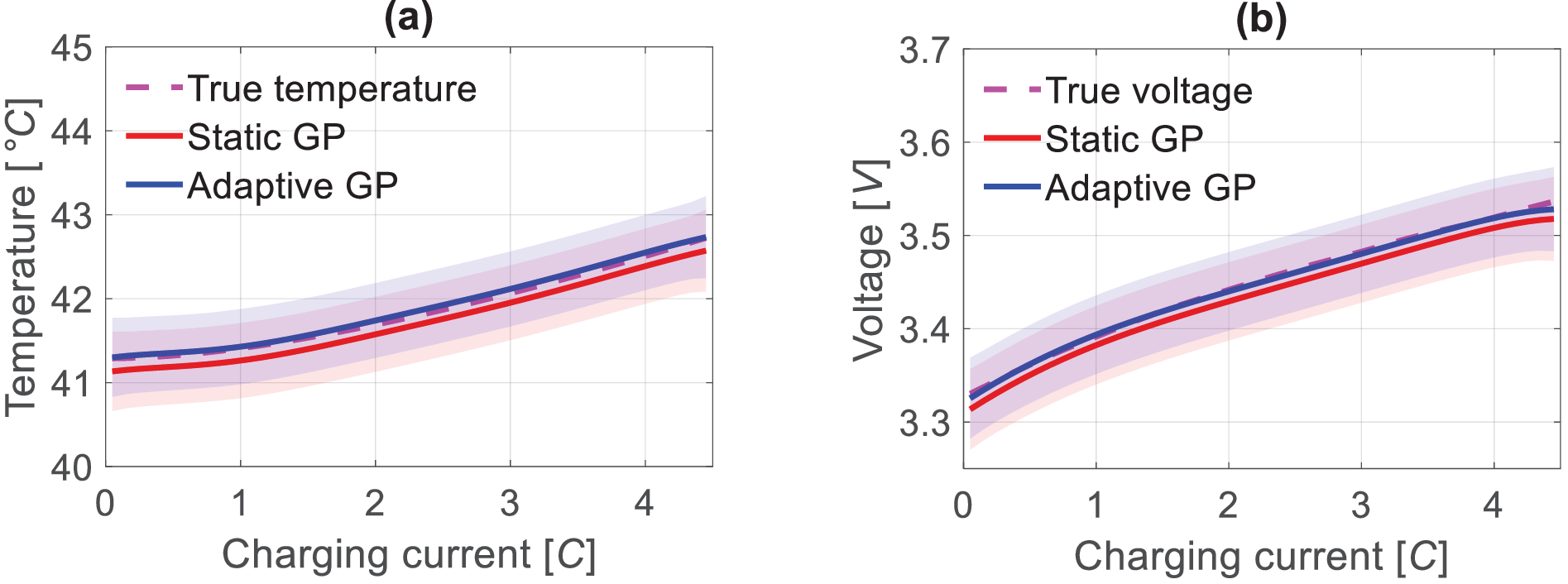}}
\caption{The predicted next-step temperature (a) and voltage (b) by static (red) and adaptive (blue) GP models against the true temperature and voltage (dashed line) under different charging currents. The GP models are trained and tested under a fixed ambient temperature $25^\degree C$. Solid lines: posterior mean; Shaded areas: $\pm$3 standard deviations.}
\label{fig:modelvalidationcase1}
\end{figure}

First, the accuracy of the static and adaptive GP models is tested in predicting the next-step temperature and voltage under a static environment, i.e., the ambient conditions are kept fixed throughout the experiments. In this work, both the static and adaptive GP models use an RBF kernel along with a white noise kernel. The RBF kernel learns the global trend between the input and output variables (described in Section \ref{subsec:fixed_gp} and \ref{subsec:adaptive_gp}), whereas white noise kernels capture the local fluctuations. The parameter values of the kernel functions are provided in Table \ref{table: Hyperparameters}. Static GPs use the data from the first $M=5$ RL episodes as the training data, and the trained static GP models are used thereafter for the entire experiment. For adaptive GPs, one trains the dynamic GPs from scratch at the beginning of each RL episode and keep updating them using the latest data in that episode. Thus, the dynamic GPs can reflect the underlying battery dynamics in real-time. 
Fig. \ref{fig:modelvalidationcase1} demonstrates the performance of both static and adaptive GP models in predicting next-step temperature and voltage of the battery under different currents to be deployed. The ambient temperature is fixed as $25 \degree C$. Fig. \ref{fig:modelvalidationcase1} shows that both static and adaptive GP models can predict the voltage and temperature accurately, and the true values of these variables fall within the uncertainty bound of the posterior means. Under fixed ambient conditions, the static GP is able to capture the dynamics of the battery well. For adaptive GP models with fixed conditions, the residuals, $\Delta T_t$ and $\Delta V_t$, are almost zero, and so are the predictions of the dynamic GP models. One critical observation is that the adaptive GPs can equally perform well as the static GPs when there are no changes in the ambient conditions. 

\begin{figure}[!t]
\centerline{\includegraphics[width= 0.9\columnwidth]{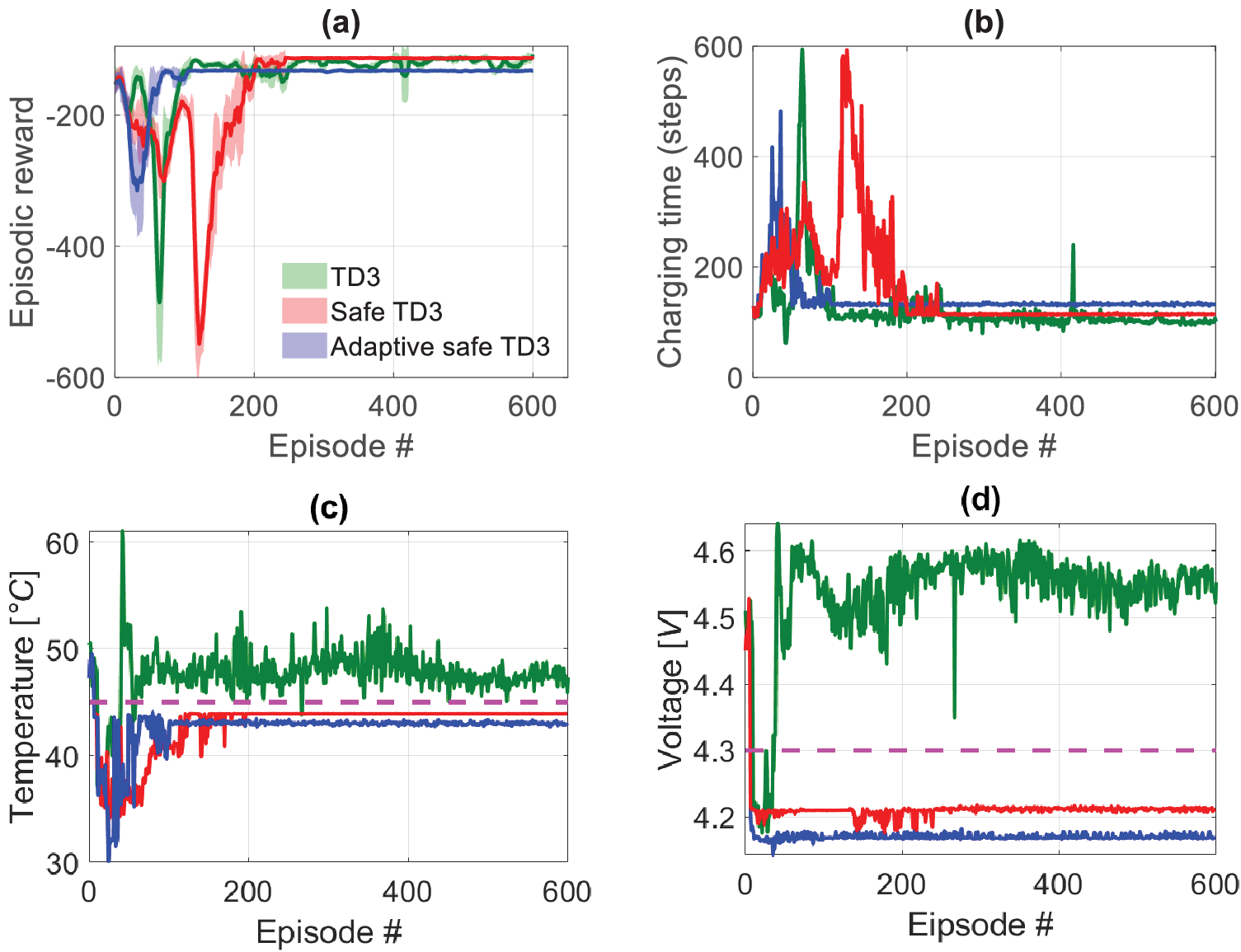}}
\caption{Training performance of traditional (green), safe (red), and adaptive safe (blue) TD3 in optimizing battery fast-charging protocols. (a) Cumulative rewards; (b) Charging time; (c) Maximum temperature; and (d) Maximum voltage,  of each training episode. Magenta dashed: the allowed upper bounds of temperature and voltage.}
\label{fig:case1a}
\end{figure}

Fig. \ref{fig:case1a} shows the training results of traditional TD3 (green), safe TD3 (red), and adaptive safe TD3 (blue) in optimizing battery fast-charging protocols. All three methods use the immediate reward $r_t$ defined in \eqref{eq:reward}, and the episodic reward can be computed as $\sum_{t=1}^{K}r_t$, where $K$ is the length of each episode. The episodic rewards in Fig. \ref{fig:case1a} (a) show that the training of traditional, safe, and adaptive safe TD3 converges to similar solutions. However, safe TD3 and adaptive safe TD3 exhibit a slower convergence rate compared to the standard TD3 method. This slowdown can be attributed to a potential discrepancy between the actions used for training the critic and actor networks within the safe TD3 algorithms (Algorithm 1). Specifically, line 30 and line 14 of Algorithm 1 employs projected actions for updating the critic network, whereas the actor network receives gradients based on the pre-projected actions (line 17). This mismatch could lead to the generation of noisy Q-values for the pre-projected actions \cite{kasaura2023benchmarking}. An attempt to address this issue is implemented by incorporating the pre-projected action $a_{t+1}$ based on the next state $s_{t+1}$ (Algorithm 1, line 13) into the target value $y_t$ used for critic network updates. While partially effective, this approach might introduce a slight delay in convergence, as observed in Fig. \ref{fig:case1a} (a). Fig. \ref{fig:case1a} (b) shows the consumed charging steps of every training episode from each method to charge the battery from 10\%-80\% SOC. Overall, the optimization makes progress as the charging time of each episode eventually decreases over episodes. Specifically, the ultimate charging time optimized by the TD3, safe TD3, and adaptive safe TD3 is shown to be 16.83, 18.83, and 22 minutes, respectively. That is, protocols optimized by safe and adaptive safe TD3 present a slower charging speed than that by traditional TD3. Fig. \ref{fig:case1a} (c)-(d) presents the maximum temperature and voltage reached during each episode, and the red dashed line indicates the allowable maximum temperature and voltage. It is seen that the protocols delivered by TD3 can lead to significant temperature and voltage violations in almost all episodes. In contrast, the protocols by safe and adaptive safe TD3 can well respect the constraints throughout the entire training. The only violations occur at the beginning when the GP models are being trained, and action projections have not been activated at that moment.

\begin{figure}[!t]
\centerline{\includegraphics[width= 0.9\columnwidth]{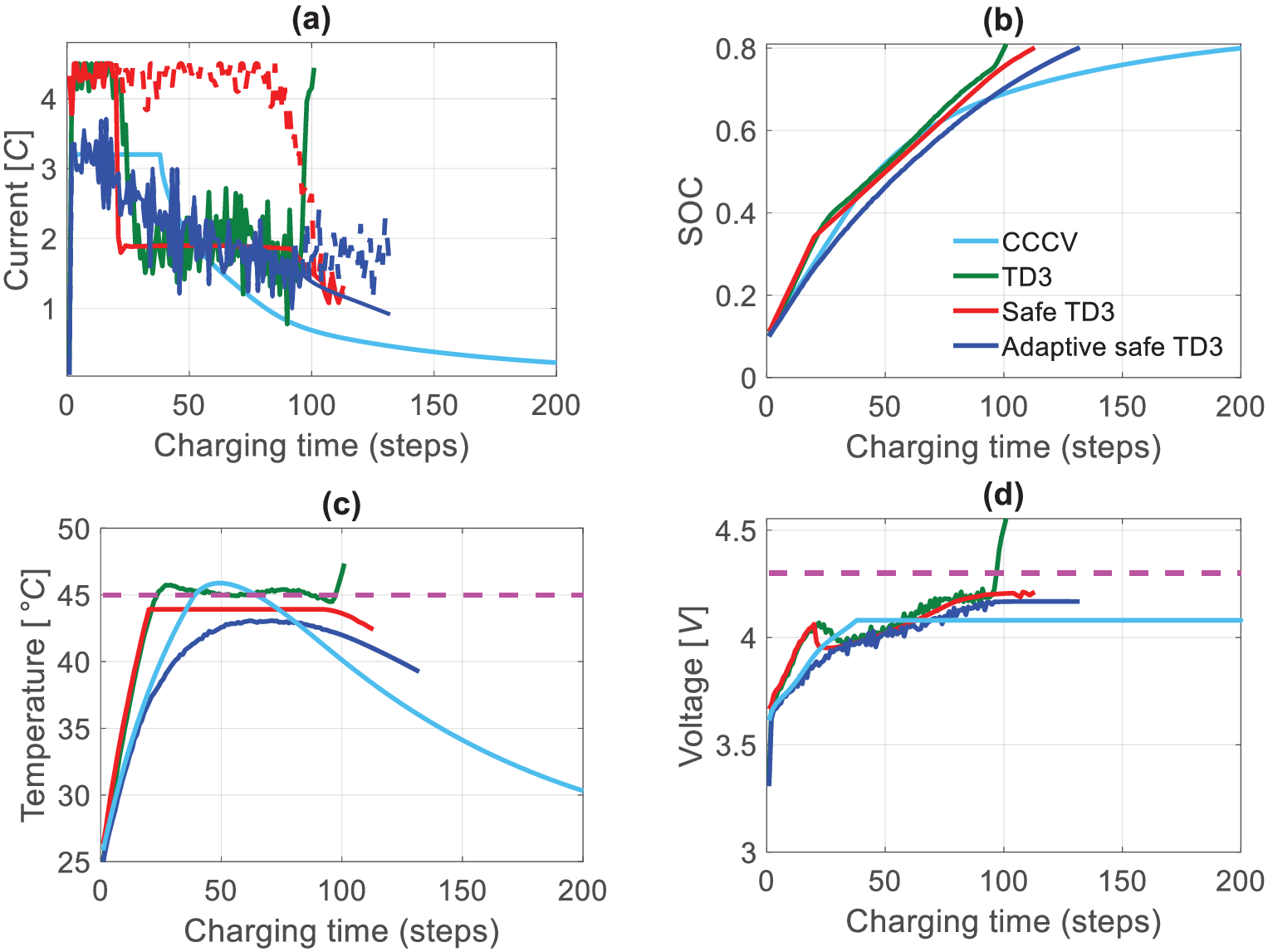}}
\caption{The (a) charging current, (b) SOC, (c) temperature, and (d) voltage profiles, obtained by deploying the optimized protocols from the traditional (green), safe (red), adaptive safe (blue) TD3, and from classical CCCV (malibu). Magenta dashed lines in (c) and (d): allowed upper bounds of the temperature and voltage. Solid lines in (a): safe current profiles after projection.}
\label{fig:case1b}
\end{figure}

Fig. \ref{fig:case1b} (a) illustrates the optimal charging profiles returned by traditional (green), safe (red), and adaptive safe (blue) TD3, as well as classical CCCV (malibu). For all protocols, a higher current is initially applied to the battery and then gradually decreases. The red and blue dashed lines are the raw currents from the actor networks, while the red and blue solid lines represent the projected currents by safe and adaptive safe TD3, respectively. Fig. \ref{fig:case1b} (a) shows that for the safe TD3 method (red), the raw charging current stays close to the upper bound for approximately 100 steps. Such a large current causes a steep rise in temperature and voltage, shown in red lines in Fig. \ref{fig:case1b} (c)-(d). As a result, the temperature of the battery approaches the temperature constraint (purple dashed) within 20 charging steps, and continuing such a large charging current onward may cause a violation. Thus, safe TD3 starts to project the current into the safety region after 20 charging steps to ensure no constraint violation. For adaptive safe TD3 (blue), the final current profile is more conservative than the safe TD3 and only requires action projection during the last 30 episodes. Fig.  \ref{fig:case1b} (b) shows that the charging protocols based on safe and adaptive safe TD3 require a slightly longer charging time than traditional TD3, however, the short charging time from traditional TD3 is at the expense of significant constraint violations (see Fig. \ref{fig:case1b} (c)-(d)). Finally, despite satisfying the operating conditions, the CCCV method takes much longer to charge the battery than all three presented RL methods.

\subsection{Safe RL-based fast charging with varying conditions}
This section tests the performance of safe and adaptive safe TD3 for battery fast-charging optimization under varying operating conditions. To this end, the ambient temperature is gradually increased from  $10^\degree C$ to $36^\degree C$ with an increment of $0.145 \degree C$ per episode starting from the 100-th episode. The diffusion-limited SEI growth model is also included to introduce battery degradation and aging \cite{xiong2020lithium}. The maximum allowable temperature $\bar{T}$ and voltage $\bar{V}$ for this case are $45^{\degree} C$ and $4.4V$, respectively, and the sampling time is 15 seconds. The architecture of the actor and critic networks is similar to the previous subsection.

\begin{figure}[!t]
\centerline{\includegraphics[width=0.9 \columnwidth, height=0.35 \columnwidth]{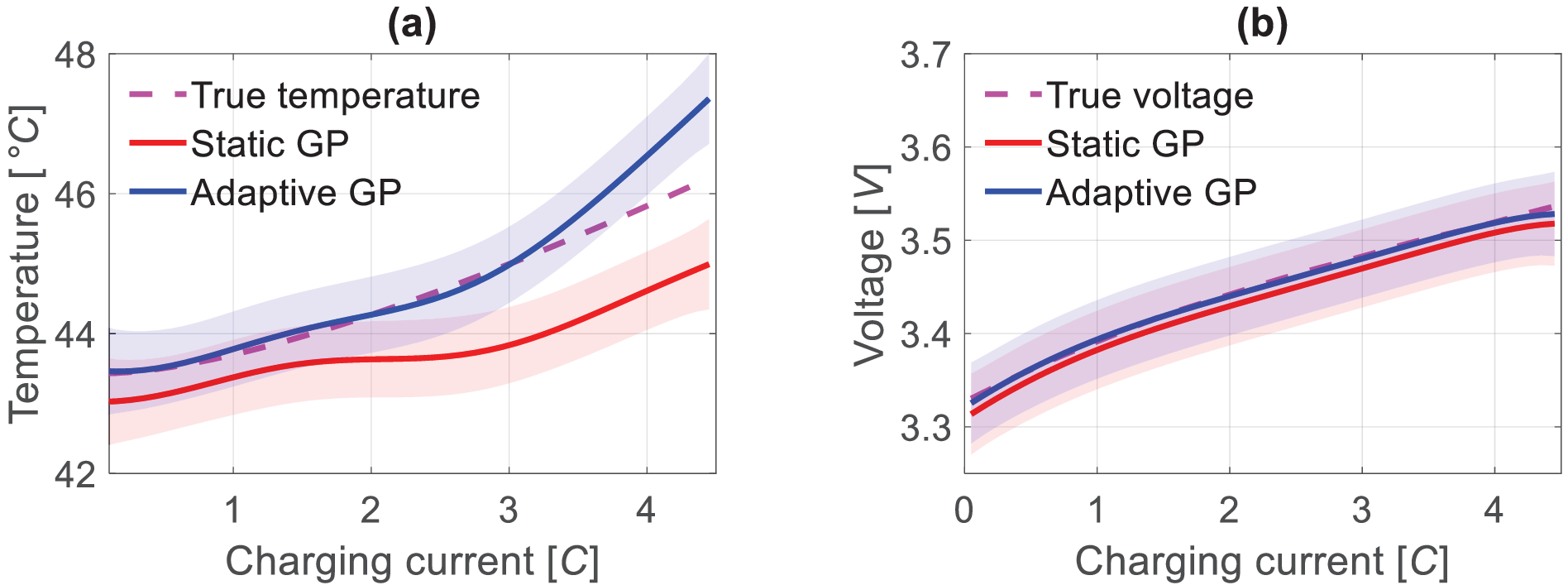}}
\caption{The predicted next-step temperature (a) and voltage (b) by static (red) and adaptive (blue) GP models against the true temperature and voltage (dashed line) under different charging currents. The GP models are trained at $10^\degree C$ and tested at $36^\degree C$ of the ambient temperature. Solid lines: posterior mean; Shaded areas: $\pm$3 standard deviations.}
\label{fig:modelvalidationcase2}
\end{figure}

This section first validates the effectiveness of static and adaptive GPs in modeling battery dynamics under varying conditions. For static GPs, the previously trained static GPs serve as the baseline (trained at $10 \degree C$), and for each episode, dynamic GPs, $\hat{f}_{\Delta}(\cdot)$ and $\hat{g}_{\Delta}(\cdot)$, are constantly updated with the online data to capture the real-time residuals. Such dynamic GPs are expected to capture the changes in battery dynamics due to variations in ambient temperatures and battery aging. Fig. \ref{fig:modelvalidationcase2} reveals that static GPs cannot capture changes in battery dynamics and fail to give accurate predictions once operating conditions change. That is, using a simple fixed constrained action space (can be defined by a fixed battery model) may not guarantee system constraints once the battery dynamics change. In contrast, adaptive GPs can well capture battery dynamics under varying conditions and battery aging. 

\begin{figure}[!t]
\centerline{\includegraphics[width= 0.9\columnwidth]{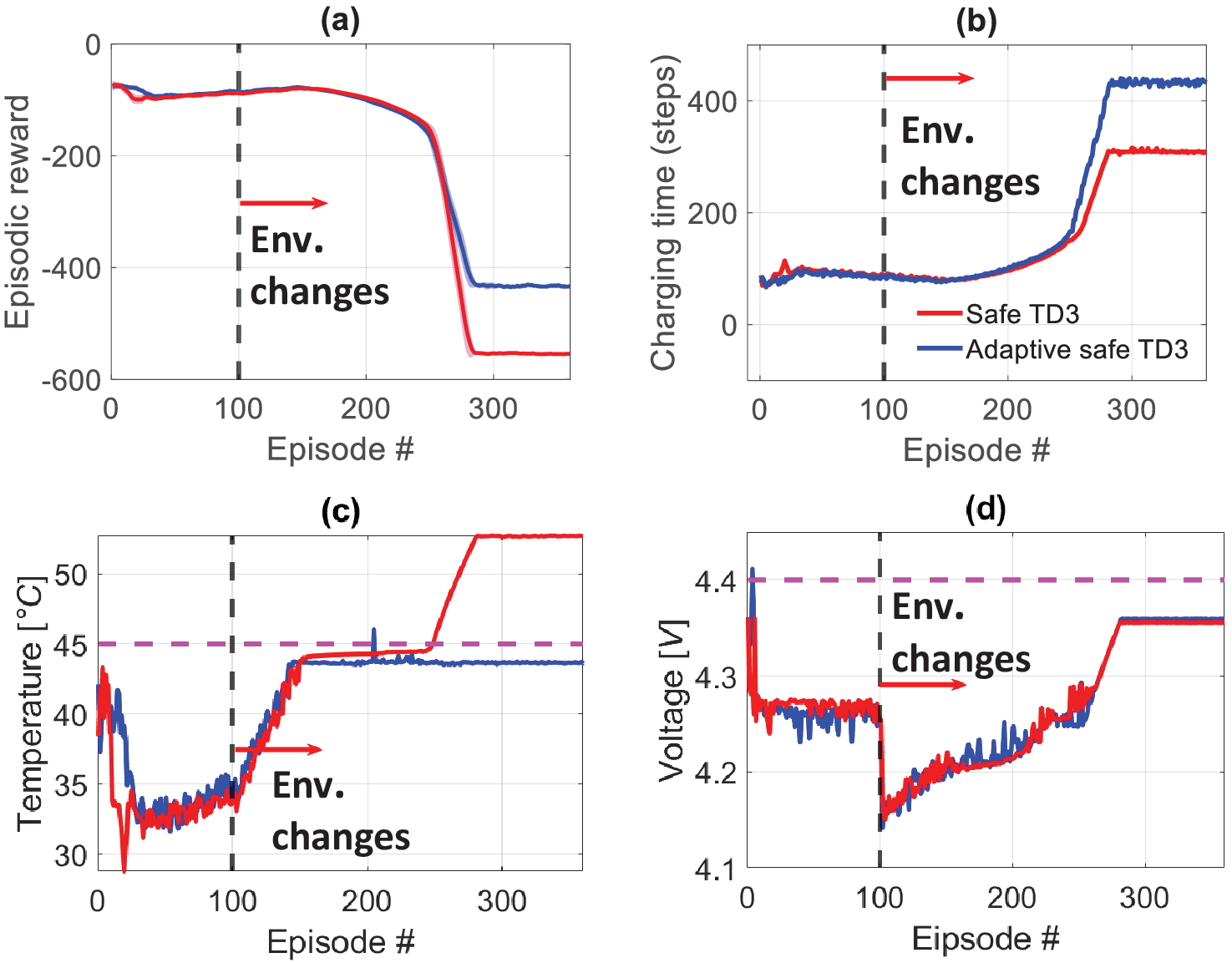}}
\caption{Training performance of safe (red) and adaptive safe (blue) TD3 for optimizing battery fast-charging protocols. (a) Cumulative rewards; (b) Charging time; (c) Maximum temperature; and (4) Maximum voltage, of each training episode. Magenta dashed: the allowed upper bounds of temperature and voltage. Vertical black dashed: the episode at which the battery operating conditions start to change.}
\label{fig:case2a}
\end{figure}	
Fig. \ref{fig:case2a} shows the training curves for the safe (red) and adaptive safe (blue) TD3 as the ambient conditions change and the battery degrades. The episodic reward curves in Fig. \ref{fig:case2a} (a) for both methods show similar behaviors. The environment change occurs at the 100-th episode due to the gradual increase in the ambient temperature from $10\degree C$ to $36 \degree C$ and the addition of the SEI growth model. After that, the rewards for both methods start to decrease since the charging current has to be less aggressive to ensure constraints under an increased ambient temperature, thereby the charging duration (see Fig. \ref{fig:case2a} (b)) is significantly enlarged that leads to lower rewards. After 250 episodes, both methods experience a sudden drop in episodic rewards. This may be due to the high ambient temperature that causes significant changes in system dynamics and rapid battery degradation. The safe TD3 cannot adjust to the new conditions and violates the temperature constraint, as in Fig. \ref{fig:case2a} (c). As a result, the temperature constraint violations lead to a large drop in episodic reward for safe TD3. In contrast, the adaptive safe TD3 can learn the latest battery dynamics online and can ensure constraints even when the environment drifts. However, this is at the price of reducing charging current and thus increasing the charging time (Fig.  \ref{fig:case2a} (b)), which also causes the rapid drop in episodic reward. 

\begin{figure}[!t]
\centerline{\includegraphics[width= 0.9\columnwidth]{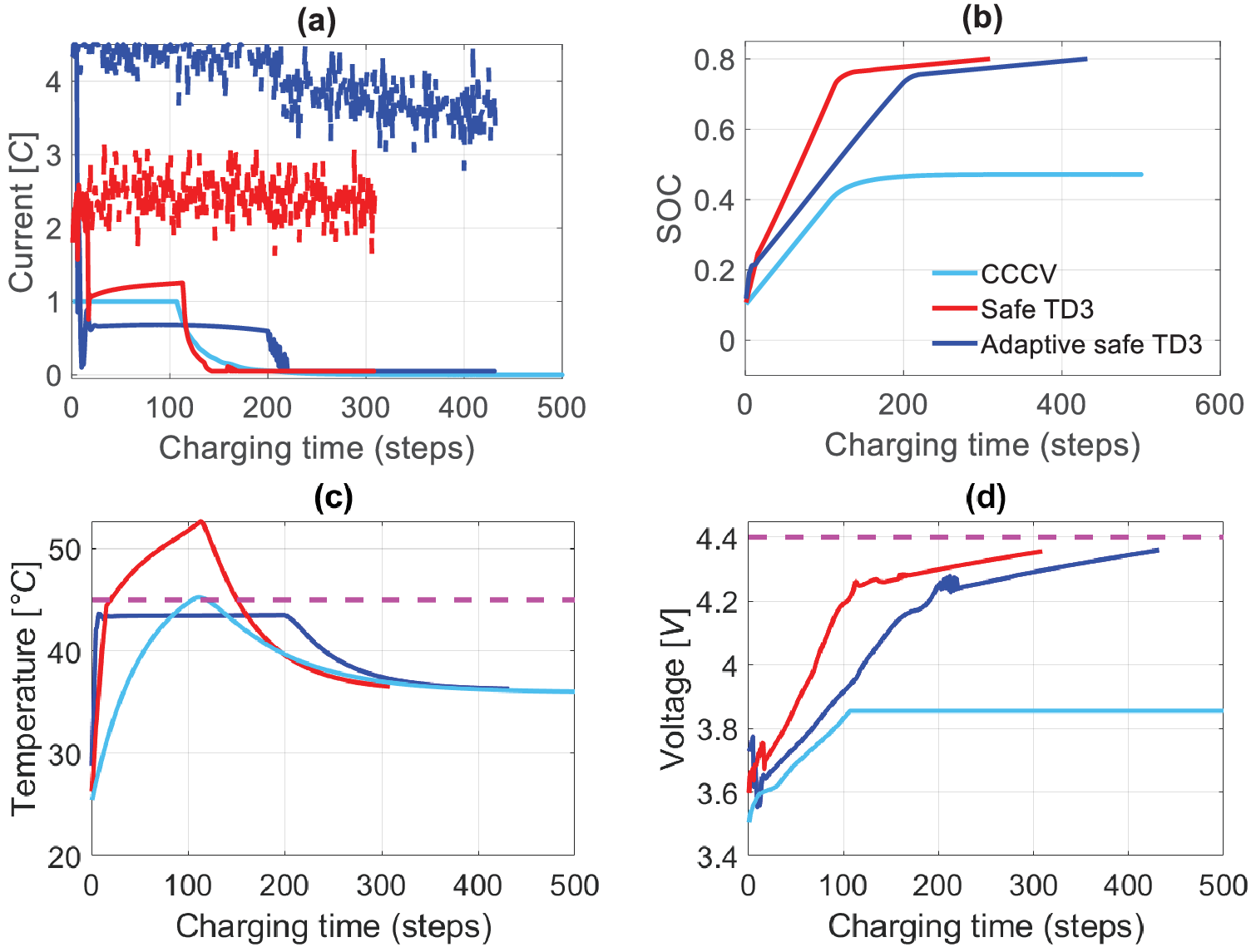}}
\caption{The (a) charging current; (b) SOC; (c) temperature; and (d) voltage profiles, obtained based on the optimized protocols from safe TD3 (red), adaptive safe TD3 (blue), and from classical CCCV (malibu). The red dashed lines in (c) and (d) indicate the allowed upper bounds of temperature and voltage. The solid lines in (a) are the safe current profiles after projection.}
\label{fig:case2b}
\end{figure}

The optimal profiles of charging currents designed by safe TD3, adaptive safe TD3, and classical CCCV are given in Fig. \ref{fig:case2b} (a). Both RL methods activate the action projection from about the 10-th step. However, as the ambient temperature increases and the battery ages, the static GP in safe TD3 cannot properly construct the new safety region, resulting in inadequate action reduction. This causes the safe TD3 to violate the temperature constraint, as shown by the red lines in Fig. \ref{fig:case2b} (c). In contrast, the adaptive safe TD3 learns the new system dynamics online and can accurately construct updated safety regions and properly project the action to respect both temperature and voltage constraints, as shown in Fig. \ref{fig:case2b} (c)-(d). However, the obtained projected actions are much more conservative than those from safe TD3, resulting in a longer charging duration, see Fig. \ref{fig:case2b} (b). Finally, for the CCCV charging protocol, the levels of constant current and constant voltage are carefully tuned to sufficiently meet the constraints. The designed charging profile is shown in malibu color of Fig. \ref{fig:case2b}. One can see that with such a design strategy, the resultant protocol cannot reach the target of 80\% SOC regardless of the charging duration (Fig. \ref{fig:case2b} (b)). In other words, the CCCV protocol is unable to tackle such harsh ambient conditions with high temperature and battery degradation. 

\begin{figure}[tbh]
\centerline{\includegraphics[width= 0.9\columnwidth]{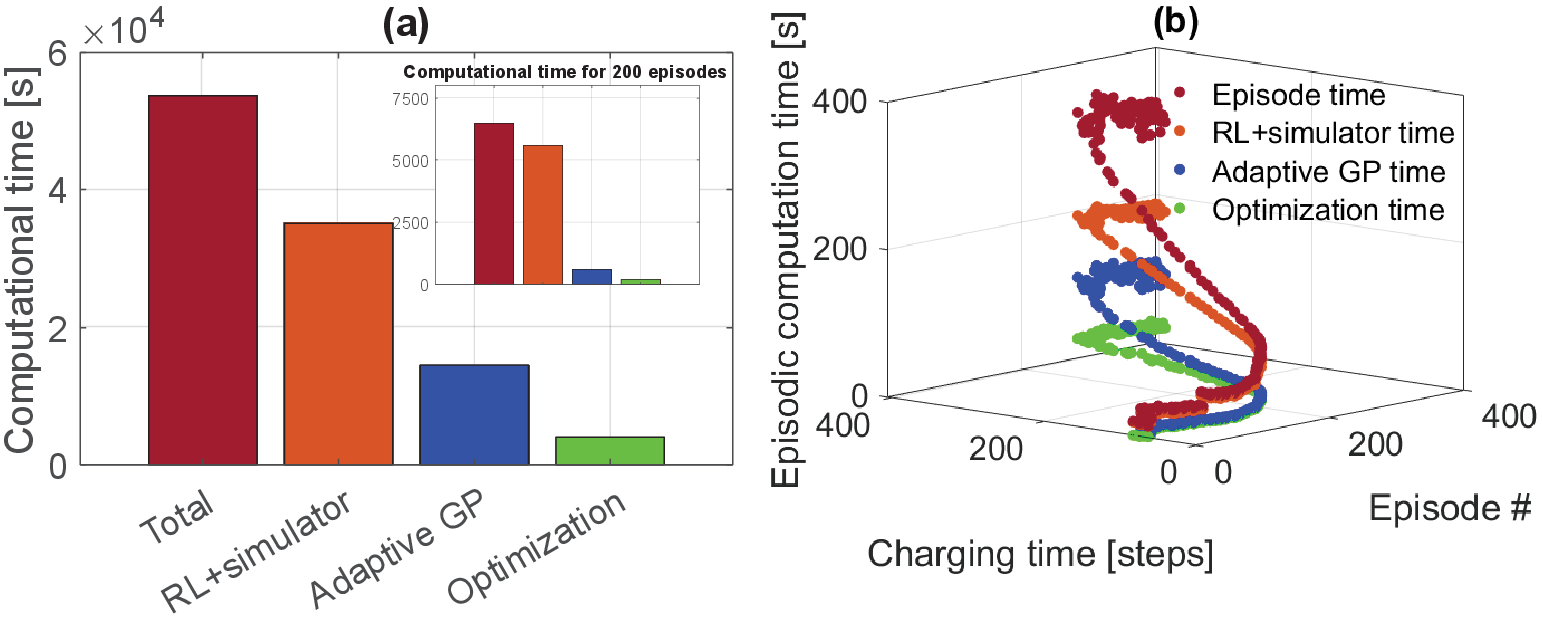}}
\caption{(a) Total computational time; and (b) Episodic computational time for each individual component of the adaptive safe TD3 method: RL, adaptive GP, and constrained optimization. The inner plot in (a) shows the computational time during the initial 200 episodes.}
\label{fig:computation time}
\end{figure}
Finally, the computational efficiency is essential for enabling the scalability, practicality, and sustainability of the proposed RL method. To this end, we determine the computation time for each element of the adaptive safe TD3 method: RL computation (action selection by the agent, execution of the action, and the evolution of the environment to  next state), adaptive GP modeling, and constrained optimization. Fig. \ref{fig:computation time} shows the total and episodic computation cost by each component. Specifically, Fig. \ref{fig:computation time} (a) reveals that the RL computation dominates the total cost, accounting for roughly 65.35\% of the total computation time. This is followed by the adaptive GP modeling (26.91\%) and constrained optimization (7.45\%). Fig. \ref{fig:computation time} (b) further delves into the computation cost by each component at every episode. When the battery is new and the ambient temperature is low (e.g., the first 200 episodes as in the inner plot in Fig. \ref{fig:computation time} (a)), the computation cost due to GP modeling (9.39\%)  and constrained optimization (3.07\%) is minimal since the favorable ambient condition allows for faster charging (few times of GP modeling and raw action as a good initial guess for the optimization). From Fig. \ref{fig:computation time} (b), as the training progresses, the increase in ambient temperature and battery degradation cause a harsh condition for battery fast-charging. The charging protocol has to be conservative to avoid constraint violations, which leads to increased computational cost for the GP modeling (more times of GP modeling) and constrained optimization (raw action is no longer close to the projected action, see Fig. \ref{fig:case2b}).

\textit{Remark}: For the proposed scheme, the RL agent will be retrained to match the latest changes in adaptive GP models by continuous online interactions with the environment. As in Fig. 6, in the presence of operating condition changes, the training of RL always goes through a transient stage during which the adaptive GP models are updated to capture latest conditions while the RL agent learns to adapt to new system (battery) dynamics and adaptive GP models.

\section{Conclusion}
\label{sec:conclusion}
This work proposed an adaptive safe RL approach (with TD3 as the exemplary RL algorithm) to optimize fast-charging protocols for batteries. Our approach introduces a safety layer after the actor network to project any unsafe actions into a safety region before being deployed to the environment. To construct such safety regions, adaptive GP models are developed as surrogates of the original black-box safety constraints. The proposed adaptive GP consists of a static GP, which captures the overall trend of battery variables, and a dynamic GP trained online with current episode data, which learns about the changes in battery dynamics caused by ambient conditions or battery aging. Extensive simulation studies were conducted based on the PyBaMM battery simulator to validate the proposed methods. Simulation results show that the adaptive GP models are effective in capturing battery dynamics when the environment drifts. Moreover, our proposed adaptive safe RL-based fast charging protocol can respect constraints with a high probability regardless of fixed or varying operating conditions. In contrast, safe RL without adaptivity can only ensure constraints under fixed conditions. Traditional TD3-based protocols cannot strictly ensure constraint satisfaction, while the CCCV-based methods are overly conservative. The proposed adaptive safe RL approach is also of theoretical and practical significance for safe RL methods to respect hard constraints in the presence of varying system dynamics. Finally, although this work mainly considers the temperature and voltage constraints of batteries, the proposed safe RL framework with adaptive GP-based surrogate modeling can be well applied to ensure satisfactions of other system constraints. Future work includes testing the proposed safe and adaptive RL methods for battery fast-charging on actual battery test beds to validate their effectiveness and robustness under real operating conditions. 

\section{Acknowledgments}
\label{sec:Acknowledgments}
M. A. Chowdhury and Saif S. S. Al-Wahaibi acknowledge the support of Distinguished Graduate Student Assistantships (DGSA) from Texas Tech University. Q. Lu acknowledges the new faculty startup funds from Texas Tech University. The authors acknowledge the support from the National Science Foundation under Grant No. 2340194. 

\bibliographystyle{aichej.bst.txt}
\bibliography{scalab}

\begin{thebibliography}{10}
\providecommand{\url}[1]{\texttt{#1}}
\providecommand{\urlprefix}{URL }

\bibitem{liu2019brief}
Liu K, Li K, Peng Q, Zhang C.
\newblock A brief review on key technologies in the battery management system
  of electric vehicles.
\newblock \emph{Frontiers of Mechanical Engineering}.
  2019;\hspace{0pt}14:47--64.

\bibitem{korkas2017adaptive}
Korkas CD, Baldi S, Yuan S, Kosmatopoulos EB.
\newblock An adaptive learning-based approach for nearly optimal dynamic
  charging of electric vehicle fleets.
\newblock \emph{IEEE Transactions on Intelligent Transportation Systems}.
  2017;\hspace{0pt}19(7):2066--2075.

\bibitem{jana2019electrochemomechanics}
Jana A, Woo SI, Vikrant K, Garc{\'\i}a RE.
\newblock Electrochemomechanics of lithium dendrite growth.
\newblock \emph{Energy \& Environmental Science}.
  2019;\hspace{0pt}12(12):3595--3607.

\bibitem{keyser2017enabling}
Keyser M, Pesaran A, Li Q, Santhanagopalan S, Smith K, Wood E, Ahmed S, Bloom
  I, Dufek E, Shirk M, et~al.
\newblock Enabling fast charging--Battery thermal considerations.
\newblock \emph{Journal of Power Sources}. 2017;\hspace{0pt}367:228--236.

\bibitem{wei2021deep}
Wei Z, Quan Z, Wu J, Li Y, Pou J, Zhong H.
\newblock {Deep deterministic policy gradient-DRL enabled
  multiphysics-constrained fast charging of lithium-ion battery}.
\newblock \emph{IEEE Transactions on Industrial Electronics}.
  2021;\hspace{0pt}69(3):2588--2598.

\bibitem{severson2019data}
Severson KA, Attia PM, Jin N, Perkins N, Jiang B, Yang Z, Chen MH, Aykol M,
  Herring PK, Fraggedakis D, et~al.
\newblock Data-driven prediction of battery cycle life before capacity
  degradation.
\newblock \emph{Nature Energy}. 2019;\hspace{0pt}4(5):383--391.

\bibitem{ahmed2017enabling}
Ahmed S, Bloom I, Jansen AN, Tanim T, Dufek EJ, Pesaran A, Burnham A, Carlson
  RB, Dias F, Hardy K, et~al.
\newblock Enabling fast charging--A battery technology gap assessment.
\newblock \emph{Journal of Power Sources}. 2017;\hspace{0pt}367:250--262.

\bibitem{thakur2023state}
Thakur AK, Sathyamurthy R, Velraj R, Saidur R, Pandey A, Ma Z, Singh P, Hazra
  SK, Sharshir SW, Prabakaran R, et~al.
\newblock A state-of-the art review on advancing battery thermal management
  systems for fast-charging.
\newblock \emph{Applied Thermal Engineering}. 2023;\hspace{0pt}226:120303.

\bibitem{bose2022study}
Bose B, Garg A, Panigrahi B, Kim J.
\newblock Study on Li-ion battery fast charging strategies: Review, challenges
  and proposed charging framework.
\newblock \emph{Journal of Energy Storage}. 2022;\hspace{0pt}55:105507.

\bibitem{notten2005boostcharging}
Notten PH, het Veld JO, Van~Beek J.
\newblock Boostcharging Li-ion batteries: A challenging new charging concept.
\newblock \emph{Journal of Power Sources}. 2005;\hspace{0pt}145(1):89--94.

\bibitem{purushothaman2006rapid}
Purushothaman B, Landau U.
\newblock Rapid charging of lithium-ion batteries using pulsed currents: A
  theoretical analysis.
\newblock \emph{Journal of The Electrochemical Society}.
  2006;\hspace{0pt}153(3):A533.

\bibitem{jiang2022fast}
Jiang B, Berliner MD, Lai K, Asinger PA, Zhao H, Herring PK, Bazant MZ, Braatz
  RD.
\newblock {Fast charging design for Lithium-ion batteries via Bayesian
  optimization}.
\newblock \emph{Applied Energy}. 2022;\hspace{0pt}307:118244.

\bibitem{klein2011optimal}
Klein R, Chaturvedi NA, Christensen J, Ahmed J, Findeisen R, Kojic A.
\newblock Optimal charging strategies in lithium-ion battery.
\newblock In: \emph{Proceedings of the 2011 american Control Conference}. IEEE.
  2011;\hspace{0pt} pp. 382--387.

\bibitem{zou2017electrochemical}
Zou C, Hu X, Wei Z, Wik T, Egardt B.
\newblock Electrochemical estimation and control for lithium-ion battery
  health-aware fast charging.
\newblock \emph{IEEE Transactions on Industrial Electronics}.
  2017;\hspace{0pt}65(8):6635--6645.

\bibitem{attia2020closed}
Attia PM, Grover A, Jin N, Severson KA, Markov TM, Liao YH, Chen MH, Cheong B,
  Perkins N, Yang Z, et~al.
\newblock Closed-loop optimization of fast-charging protocols for batteries
  with machine learning.
\newblock \emph{Nature}. 2020;\hspace{0pt}578(7795):397--402.

\bibitem{pozzi2022deep}
Pozzi A, Moura S, Toti D.
\newblock Deep Learning-Based Predictive Control for the Optimal Charging of a
  Lithium-Ion Battery with Electrochemical Dynamics.
\newblock In: \emph{2022 IEEE Conference on Control Technology and Applications
  (CCTA)}. IEEE. 2022;\hspace{0pt} pp. 785--790.

\bibitem{hao2023adaptive}
Hao Y, Lu Q, Wang X, Jiang B.
\newblock Adaptive Model-Based Reinforcement Learning for Fast Charging
  Optimization of Lithium-Ion Batteries.
\newblock \emph{IEEE Transactions on Industrial Informatics}.
  2023;\hspace{0pt}.

\bibitem{sutton2018reinforcement}
Sutton RS, Barto AG.
\newblock \emph{Reinforcement Learning: An Introduction}.
\newblock MIT press. 2018.

\bibitem{park2020reinforcement}
Park S, Pozzi A, Whitmeyer M, Perez H, Joe WT, Raimondo DM, Moura S.
\newblock Reinforcement learning-based fast charging control strategy for
  li-ion batteries.
\newblock In: \emph{2020 IEEE Conference on Control Technology and Applications
  (CCTA)}. IEEE. 2020;\hspace{0pt} pp. 100--107.

\bibitem{park2022deep}
Park S, Pozzi A, Whitmeyer M, Perez H, Kandel A, Kim G, Choi Y, Joe WT,
  Raimondo DM, Moura S.
\newblock A deep reinforcement learning framework for fast charging of li-ion
  batteries.
\newblock \emph{IEEE Transactions on Transportation Electrification}.
  2022;\hspace{0pt}8(2):2770--2784.

\bibitem{chang2020control}
Chang F, Chen T, Su W, Alsafasfeh Q.
\newblock Control of battery charging based on reinforcement learning and long
  short-term memory networks.
\newblock \emph{Computers \& Electrical Engineering}.
  2020;\hspace{0pt}85:106670.

\bibitem{kim2023optimal}
Kim M, Lim J, Ham KS, Kim T.
\newblock Optimal charging method for effective Li-ion battery life extension
  based on reinforcement learning.
\newblock In: \emph{Proceedings of the 38th ACM/SIGAPP Symposium on Applied
  Computing}. 2023;\hspace{0pt} pp. 1659--1661.

\bibitem{yang2023balancing}
Yang Y, He J, Chen C, Wei J.
\newblock Balancing Awareness Fast Charging Control for Lithium-Ion Battery
  Pack Using Deep Reinforcement Learning.
\newblock \emph{IEEE Transactions on Industrial Electronics}.
  2023;\hspace{0pt}.

\bibitem{elouazzani4218801smart}
Elouazzani H, Elhassani I, Barka N, Masrour T.
\newblock Smart Adaptive Multi Stage Constant Current Fast Charging for Lithium
  Ions Batteries Based on Deep Reinforcement Learning.
\newblock \emph{Available at SSRN 4218801};\hspace{0pt}.

\bibitem{fujimoto2018addressing}
Fujimoto S, Hoof H, Meger D.
\newblock Addressing function approximation error in actor-critic methods.
\newblock In: \emph{International Conference on Machine Learning}. PMLR.
  2018;\hspace{0pt} pp. 1587--1596.

\bibitem{lillicrap2015continuous}
Lillicrap TP, Hunt JJ, Pritzel A, Heess N, Erez T, Tassa Y, Silver D, Wierstra
  D.
\newblock Continuous control with deep reinforcement learning.
\newblock \emph{arXiv preprint arXiv:150902971}. 2015;\hspace{0pt}.

\bibitem{wang2020intuitive}
Wang J.
\newblock {An intuitive tutorial to Gaussian processes regression}.
\newblock \emph{arXiv preprint arXiv:200910862}. 2020;\hspace{0pt}.

\bibitem{richardson2017gaussian}
Richardson RR, Osborne MA, Howey DA.
\newblock Gaussian process regression for forecasting battery state of health.
\newblock \emph{Journal of Power Sources}. 2017;\hspace{0pt}357:209--219.

\bibitem{haarnoja2018soft}
Haarnoja T, Zhou A, Hartikainen K, Tucker G, Ha S, Tan J, Kumar V, Zhu H, Gupta
  A, Abbeel P, et~al.
\newblock Soft actor-critic algorithms and applications.
\newblock \emph{arXiv preprint arXiv:181205905}. 2018;\hspace{0pt}.

\bibitem{levine2020offline}
Levine S, Kumar A, Tucker G, Fu J.
\newblock Offline reinforcement learning: Tutorial, review, and perspectives on
  open problems.
\newblock \emph{arXiv preprint arXiv:200501643}. 2020;\hspace{0pt}.

\bibitem{xiong2020lithium}
Xiong R, Pan Y, Shen W, Li H, Sun F.
\newblock {Lithium-ion battery aging mechanisms and diagnosis method for
  automotive applications: Recent advances and perspectives}.
\newblock \emph{Renewable and Sustainable Energy Reviews}.
  2020;\hspace{0pt}131:110048.

\bibitem{sulzer2021python}
Sulzer V, Marquis SG, Timms R, Robinson M, Chapman SJ.
\newblock {Python battery mathematical modelling (PyBaMM)}.
\newblock \emph{Journal of Open Research Software}. 2021;\hspace{0pt}9(1).

\bibitem{kasaura2023benchmarking}
Kasaura K, Miura S, Kozuno T, Yonetani R, Hoshino K, Hosoe Y.
\newblock Benchmarking actor-critic deep reinforcement learning algorithms for
  robotics control with action constraints.
\newblock \emph{IEEE Robotics and Automation Letters}.
  2023;\hspace{0pt}8:4449--4456.

\end{thebibliography}
\end{document}